\input{aipcheck}

\documentclass[
    ,final            
  ]
  {aipproc}
\usepackage{multirow}
\usepackage{amssymb}  
\usepackage{booktabs}
\usepackage{subfigure}  
  
\layoutstyle{8x11single}
\def\slashchar#1{\setbox0=\hbox{$#1$}
   \dimen0=\wd0 \setbox1=\hbox{/} \dimen1=\wd1
   \ifdim\dimen0>\dimen1 \rlap{\hbox to \dimen0{\hfil/\hfil}} #1
   \else  \rlap{\hbox to \dimen1{\hfil$#1$\hfil}} / \fi}

\begin{document}

\title{Weak Quasielastic Production of Hyperons and Threshold Production of Two Pions.}

\classification{<Replace this text with PACS numbers; choose from this list:
                \texttt{http://www.aip..org/pacs/index.html}>}

\author{S. K. Singh}{
  address={Department of Physics, Aligarh Muslim University, Aligarh-202 002, India}
}

\author{M. Sajjad Athar}{
  address={Department of Physics, Aligarh Muslim University, Aligarh-202 002, India}
}

\author{M. Rafi Alam}{
  address={Department of Physics, Aligarh Muslim University, Aligarh-202 002, India}
}

\author{Shikha Chauhan}{
  address={Department of Physics, Aligarh Muslim University, Aligarh-202 002, India}
}

\author{E. Hern\'andez}{
  address={Grupo de F\'\i sica Nuclear,
Departamento de F\'\i sica Fundamental e IUFFyM, Universidad de
Salamanca, E-37008 Salamanca, Spain.}
}

\author{J. Nieves}{
 address={Departamento de F\'\i sica Te\'orica and Instituto de F\'isica Corpuscular, Centro Mixto
Universidad de Valencia-CSIC, E-46071 Valencia, Spain}
}

\author{M. Valverde}{
 address={Departamento de F\'\i sica Te\'orica and Instituto de F\'isica Corpuscular, Centro Mixto
Universidad de Valencia-CSIC, E-46071 Valencia, Spain}
}

\author{M. J. Vicente Vacas}{
  address={Departamento de F\'\i sica Te\'orica and Instituto de F\'isica Corpuscular, Centro Mixto
Universidad de Valencia-CSIC, E-46071 Valencia, Spain}
}

\classification{13.15+g, 13.75.Ev, 14.20.Jn, 21.60.Jz, 25.30.Pt}

\begin{abstract}
We have studied quasielastic charged current hyperon production induced by $\bar\nu_\mu$ on free nucleon
and the nucleons bound inside the nucleus and the results are presented for several nuclear targets like 
$^{40}Ar$, $^{56}Fe$ and $^{208}Pb$. The hyperon-nucleon transition form
factors are determined from neutrino-nucleon scattering and semileptonic decays
of neutron and hyperons using SU(3) symmetry.  The nuclear medium
effects(NME) due to Fermi motion and final state interaction(FSI) effect due to 
hyperon-nucleon scattering have been taken into account.
Also we have studied two pion production at threshold induced by neutrinos off nucleon targets. The
contribution of nucleon, pion, and contact terms are calculated using Lagrangian given 
by nonlinear $\sigma$ model. The contribution
of the Roper resonance has also been taken into account. The numerical
results for the cross sections are presented and compared with the experimental results 
from ANL and BNL. 

\end{abstract}

\maketitle

\section{Introduction}

The neutrino oscillation experiments being done using accelerator (anti)neutrino 
beams in the energy region of few GeV are also providing cross section measurements 
of various reactions induced by neutrinos and antineutrinos on nuclear targets 
which are needed for validating various Monte Carlo neutrino event generators. 
The reported cross sections are available mainly for elastic, quasielastic and 
single pion production processes on $^{12}C$ and $^{16}O$ nuclei. The availability of high 
intensity (anti)neutrino
beams in present generation neutrino oscillation experiments has opened up 
the possibility of cross section measurements
of other processes like production of hyperons, strange baryons and multiple pions in the
energy region of a few GeV
in the experiments being done at MINER$\nu$A~\cite{Solomey:2005rs}, MINOS~\cite{Evans:2013pka},
NO$\nu$A~\cite{Ayres:2004js}, LBNE~\cite{Adams:2013qkq}, 
T2K~\cite{Abe:2011ks} and MiniBooNE~\cite{AguilarArevalo:2007it}. In the lower energy region relevant 
for T2K~\cite{Abe:2011ks} and MicroBooNE~\cite{Chen:2007ae} experiments, any observation of 
these processes will help to understand the reaction dynamics
of the threshold production of strange mesons and two pions. In view of this scenario, 
some calculations have been done
recently for the production of hyperons~\cite{Alam:2013}, kaons~\cite{RafiAlam:2010kf,Alam:2012zz} 
and two pions~\cite{Hernandez:2007ej}.

In this work we present a brief overview of the quasieleastic production of hyperons induced by antineutrinos 
on nucleons and nuclei and charged current two pion production induced by neutrino from nucleon.
First, we are going to present the formalism, results and discussion for the quasieleastic 
production of hyperons from nucleons and nuclear targets 
followed by the formalism of two pion
production from nucleons and their results and discussion. 

\section{Quasielastic Production of Hyperons}
\subsection{Formalism}

The quasieleastic production of hyperons induced by antineutrinos has been studied in the past using Cabibbo theory
with SU(3) symmetry~\cite{Singh:2006xp} as well as in quark models~\cite{Wu:2013kla}. 
In addition to providing neutrino-nucleus cross section 
to be used in modeling the MC neutrino event generators, the differential cross sections also provide an opportunity
to study nucleon-hyperon transition form factors at high $Q^2$, which are available only at low $Q^2$ through 
the analysis of Semileptonic Hyperon Decays(SHD)~\cite{Cabibbo:2003cu}.
An independent determination of these form factors help to test
various assumptions made in the analysis of these processes like SU(3) symmetry, G invariance, absence of Flavor
Changing Neutral Currents(FCNC), $\Delta Q = \Delta S$ rule, Conserved Vector Current(CVC) and Partial Conservation
of Axial Current(PCAC) hypothesis in the Standard Model(SM) when it is extended to the strangeness sector. 
Since these hyperons decay  primarily  through pionic decay modes, they also contribute to the pion 
production processes induced by antineutrinos, which is dominated by the pion production 
through $\Delta$-excitation. In the low energy region where $\Delta$-excitation is inhibited by threshold effects, 
the Cabibbo suppressed pion production through hyperons$(\Lambda \; \rm{and}  \; \Sigma)$ production may be important. 

We consider the following processes
\begin{eqnarray}\label{hyp-rec}
{\bar\nu_\mu}(k) + p(p)&\rightarrow \mu^+(k^\prime) + \Lambda(p^\prime) \nonumber \\ 
{\bar\nu_\mu}(k) + p(p)&\rightarrow \mu^+(k^\prime) + \Sigma^0(p^\prime)\nonumber \\ 
{\bar\nu_\mu}(k) + n(p)&\rightarrow \mu^+(k^\prime) + \Sigma^-(p^\prime) ,
\end{eqnarray} 

where $k(k^\prime)$ and $p(p^\prime)$  are the momenta of intial(final) lepton and nucleon. 
The differential scattering cross section is given by,
\begin{equation}
\label{crosv.eq}
d\sigma=\frac{1}{(2\pi)^2}\frac{1}{4E_{\bar \nu} M}\delta^4(k+p-k^\prime-p^\prime)
\frac{d^3k^\prime}{2E_{k^\prime}}\frac{d^3p^\prime}{2E_{p^\prime}}\sum \bar{\sum} |{\cal{M}}|^2 ,
\end{equation}
with
\begin{eqnarray}
\quad {\cal{M}} &=& \frac{G_F}{\sqrt{2}}\sin\theta_c l^{\mu}~J_\mu \\ 
l^\mu &=& \bar v(k^\prime) \gamma^\mu (1+\gamma_5) v(k) \\
J_{\mu} &=& \langle Y(p^{\prime})|V_{\mu} - A_{\mu}|N(p) \rangle ,
\end{eqnarray}
where
\begin{eqnarray}\label{vec}
 \langle Y(p^{\prime})|V_{\mu}|N(p) \rangle &=& {\bar{u}_Y}(p^\prime)\left[\gamma_\mu f_1(q^2)+i\sigma_{\mu\nu} 
\frac{q^\nu}{M+M_Y} f_2(q^2) +
\frac{f_3(q^2)}{M + M_Y} q_\mu \right]u_N(p) \\
 \langle Y(p^{\prime})|A_{\mu}|N(p) \rangle &=& {\bar{u}_Y}(p^\prime)\left[\gamma_\mu \gamma_5 g_1(q^2) + 
 i \sigma_{\mu\nu}\gamma_5 \frac{q^\nu}{M+M_Y} g_2(q^2) +
 \frac{g_3(q^2)} {M + M_Y} q_\mu \gamma_5 \right]u_N(p),
\end{eqnarray}

The form factors $f_i(q^2)$ and $g_i(q^2)$ are determined using  T invariance, G invariance, SU(3) symmetry 
(symmetry properties of weak currents) like CVC and PCAC hypothesis. 
These symmetry considerations yield~\cite{Singh:2006xp}:
\begin{enumerate}
 \item $f_i(q^2)$ and $g_i(q^2)$ as real quantities
 \item $f_3(q^2) = g_2(q^2) = 0 $
 \item $g_3(q^2) = \frac{m_\pi}{m_\pi^2 - q^2} g_1(q^2) $ 
 \item $f_{i=1,2}^{p\rightarrow \Sigma^0} \left( g_{i=1,2}^{p\rightarrow \Sigma^0}\right)  = 
  \frac{1}{\sqrt2} \;\; f_{i=1,2}^{p\rightarrow \Sigma^-} \left( g_{i=1,2}^{p\rightarrow \Sigma^-}\right)  $
 \item The form factors, $f_{1,2} (q^2) = \frac{-1}{\sqrt2} (f_{1,2}^p(q^2) + 2 f_{1,2}^n(q^2)) $ and 
    $f_{1,2} (q^2) = -\sqrt{\frac32} f_{1,2}^p(q^2)$ respectively for transitions $p \rightarrow \Sigma^0 $ and
 $p \rightarrow \Lambda $. Similarly  $g_1(q^2)$ is $\frac{1}{\sqrt2}\frac{D-F}{D+F} g_A(q^2)$ for 
 $p \rightarrow \Sigma^0 $ and is $-\frac{D+3 F}{\sqrt{6} (D+F)} g_A(q^2)$  for $p \rightarrow \Lambda $ transitions.

We have used $D=0.804$ and $F=0.463$ and the following $q^2$ dependence of the 
electroproduction and weak form factors~\cite{Alam:2014bya}:

\begin{eqnarray}\label{Vec_FF}
 f_{1}^{p, n}(q^2) = 
 \frac{1}{1 - \frac{q^2}{4M^2}} \left[ G_{E}^{p, n}(q^2) - \frac{q^2}{4M^2} G_{M}^{p, n}(q^2)\right],  \quad  \qquad  
 f_{2}^{p, n}(q^2) = \frac{1}{1 - \frac{q^2}{4M^2}} \left[ G_{M}^{p, n}(q^2) - G_{E}^{p, n}(q^2)\right] .
\end{eqnarray}
The Sach's form factors $G_{E}^{p, n}(q^2)$ and $G_{M}^{p, n}(q^2)$ are parameterized as 
\begin{eqnarray}
 G_{E}^{p}(q^2)  &=&  \left( 1 - \frac{q^2}{M_{V}^{2}}\right)^{-2},  \quad  \qquad  
 G_{M}^{p}(q^2)  = (1 + \mu_{p}) G_{E}^{p}(q^2),   \nonumber\\
 G_{M}^{n}(q^2) &=& \mu_{n} G_{E}^{p}(q^2), \quad  \qquad
 G_{E}^{n}(q^2) = \frac{q^2}{4M^2} \mu_{n} G_{E}^{p}(q^2) \xi_{n};
\end{eqnarray}
The numerical values of various parameters are taken as
\begin{eqnarray}
 \xi_{n} &=& \frac{1}{1 - \lambda_{n} \frac{q^2}{4M^2}}, \quad  \qquad 
 \mu_{p} = 1.792847, \nonumber\\
 \mu_{n} &=& -1.913043, \quad  \qquad 
 M_{V} = 0.84 GeV  \;\; \rm{and} \;\;  \lambda_{n} = 5.6
\end{eqnarray}
For $g_A(q^2)$ a dipole form has been taken i.e.  $g_A(q^2) = g_{A}(0) \left(1 - \frac{q^2}{M_{A}^{2}}\right)^{-2} $ 
with $g_{A}(0) =1.267$ and the  axial dipole mass $M_{A} = 1$GeV.  
 
\end{enumerate}

\subsection{Nuclear Medium Effects and Final State Interaction Effects}

When these reactions take place on bound nucleons in nuclear medium, Fermi motion and 
Pauli Blocking effects of nucleons are to be considered. 
In the final state after hyperons are produced, they may undergo strong interaction 
scattering processes through charge exchange( $\Sigma^- p \rightarrow \Lambda  n $, 
$\Lambda p \rightarrow \Sigma^+ n$, etc.) and inelastic ($\Lambda N \rightarrow \Sigma^0 N $,
$Y N \rightarrow Y^\prime N^\prime $)
reactions like  changing the relative yield of $\Sigma^0,\; \Sigma^- $ and $\Lambda$ 
produced in the initial reactions shown in Eq.~\ref{hyp-rec}. 
In a special case, $\Sigma^+$ will appear as a result of final state interaction which are initially not produced 
through $\bar \nu_\mu N \rightarrow \mu^+ N$ reaction due to $\Delta Q = \Delta S$ rule.

The nuclear medium effects are calculated in a relativistic Fermi Gas model 
using local density approximation and the nuclear cross section is written as 
\begin{equation}\label{diffnuc}
\frac{d\sigma}{d\Omega_ldE_l}=2{\int d^3r \int \frac{d^3p}{{(2\pi)}^3}n_N(p,r)
\left[\frac{d\sigma}{d\Omega_ldE_l}\right]_{free}}
\end{equation}
where $n_N(p,r)$ is local occupation number of the initial 
nucleon of momentum $p$ and is 1 for $p < p_{F_N}$ and 0 otherwise with 
\begin{equation}
{p_F}_n={[3\pi^2\rho_n(r)]}^{1/3} \qquad \qquad \rm{for} \qquad N=n,p
\end{equation}
The final state interaction of hyperon-nucleon system is calculated 
in a Monte Carlo simulation approach. 
In this approach an initial hyperon produced at a position ${\bf r}$ within the nucleus which interacts with 
a nucleon to produce a new hyperon-nucleon state $f=Y_f N_f$, within a distance $l$ with 
probability $P_Y \, dl$ where $P_Y$ is the probability per unit length given by
\begin{equation}
 P_Y=\sigma_{Y+n \rightarrow f}(E)~\rho_{n}(r)~+~\sigma_{Y+p \rightarrow
f}(E)~\rho_{p}(r),
\end{equation}
where $\rho_{n}(r)(\rho_{p}(r))$ is the local density of neutron(proton) in the
nucleus and $\sigma$ is the  cross section for $YN\rightarrow f$ process. 
Now a particular channel $f$ is selected. For the selected channel $f$, a hyepron-nucleon 
state is chosen by randomly selecting the momentum of initial nucleon consistent with Pauli blocking. 
A random scattering angle is generated using isotropic cross section for the hyperon-nucleon scattering
cross section $\sigma$, which then determines momentum(energy) of final hyperon and nucleon. 
If the momentum of final nucleon is above Fermi momentum, a new final state of $YN$ system is obtained. 
This process is continued until the hyperon gets out of the nucleus. 
All the channels i.e. $Y+n \rightarrow Y_f + N_f$ and $Y+p \rightarrow Y_f + N_f$ leading 
to a final state $f$ are considered. 
These hyperons decay into pions through the processes. 
\begin{eqnarray}
 \Sigma^0 (\Lambda^0)   \rightarrow   p \pi^-, n \pi^0 , \qquad 
 \Sigma^+   \rightarrow  p \pi^0, n \pi^+ , \qquad  
 \Sigma^-  \rightarrow   n \pi^- 
\end{eqnarray}
In the final state, a $\mu^+$ will be accompanied either by a $\pi^0$ or $\pi^-$ 
and rarely by a $\pi^+ $  which will be produced as a result of the final state interaction in the nucleus. 
Note that it can also be produced as final state interaction effect when 
$\pi^0 p \rightarrow \pi^+ n$ reaction takes place.

\subsection{Results}

 \begin{figure}
\includegraphics[height=0.26\textheight,width=0.7\textwidth]{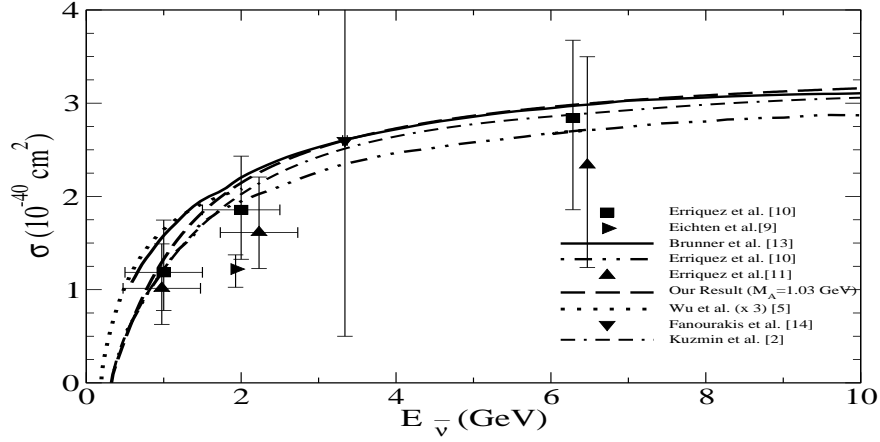}
\caption{$\sigma$ vs $E_{\bar\nu_\mu}$, for ${\bar\nu_\mu} + p \rightarrow \mu^+ 
+ \Lambda$ process. Experimental results (triangle right~\cite{Eichten:1972bb}, square~\cite{Erriquez:1978pg},
triangle up~\cite{Erriquez:1977tr}, circle~\cite{Brunner:1989kw}), 
triangle down($\sigma = 2.6^{+5.9}_{-2.1} \times 10^{-40} cm^2$)~\cite{Fanourakis:1980si}
are shown with error bars. Theoretical curves are of Erriquez et al.~\cite{Erriquez:1978pg}(dashed-double dotted line), Brunner et al.~\cite{Brunner:1989kw}(dashed line), 
 and Kuzmin and Naumov~\cite{Kuzmin:2008zz}(double dashed-dotted line) obtained 
 using Cabibbo theory  with axial vector dipole mass as 1 GeV, 1.1 GeV and 0.999GeV respectively, while
 the results of Wu et al.~\cite{Wu:2013kla}(dotted line) and Finjord and 
Ravndal~\cite{Finjord:1975zy}(dashed dotted line) are obtained using quark model. 
The results of present calculation are 
shown with solid line. Notice that we have multiplied the results of 
Wu et al.~\cite{Wu:2013kla} by 3 to plot on the same scale.}
\label{fg:xsec_comp}
\end{figure}

\begin{figure}[h]
\includegraphics[height=0.26\textheight,width=0.7\textwidth]{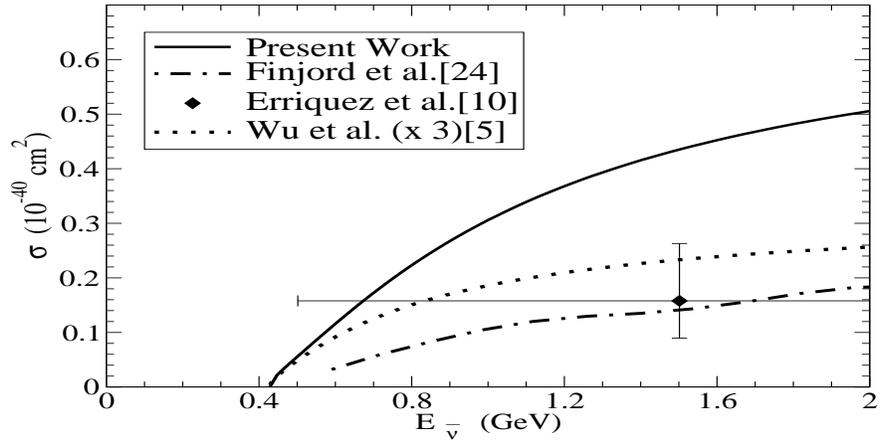}
\caption{$\sigma$ vs $E_{\bar\nu_\mu}$, for $\bar\nu_\mu + p \rightarrow \mu^+ + \Sigma^0$ process. 
Experimental points is taken from~\cite{Erriquez:1978pg}.
Present results are shown with solid line. Also the results of Wu et al.~\cite{Wu:2013kla}(dotted line) 
and  Finjord and Ravndal~\cite{Finjord:1975zy}(dashed dotted line) have been presented. 
Notice that we have multiplied the results of Wu et al.~\cite{Wu:2013kla} by 3 to plot on the same scale.}
\label{fg:xsec_p_mu_Sig}     
\end{figure}

\begin{figure}
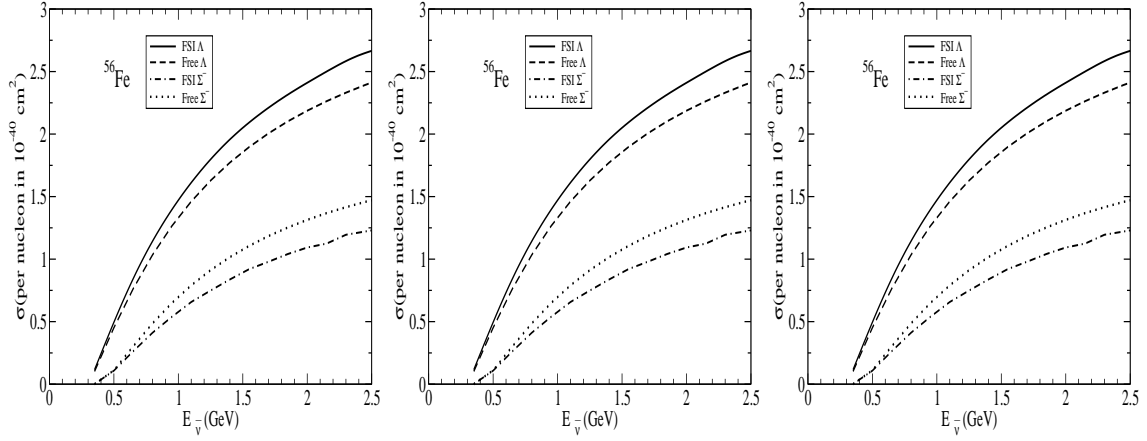

 \includegraphics[height=0.26\textheight,width=0.3\textwidth]{xsec_fe.eps}
  \includegraphics[height=0.26\textheight,width=0.3\textwidth]{xsec_fe.eps}
 \includegraphics[height=0.26\textheight,width=0.3\textwidth]{xsec_fe.eps}
\caption{$\sigma$ vs $E_{\bar\nu_\mu}$ for $^{40}Ar, ^{56}Fe$ and $^{208}Pb$ nuclei. }
\label{fg:xsec_fe_ar_pb}     
\end{figure}

In Fig.~\ref{fg:xsec_comp}, we have presented the results for the total cross section for
$\bar \nu_\mu + p \rightarrow \mu^+ + \Lambda $ reaction for free nucleons and compared them with the 
experimental results~\cite{Eichten:1972bb,Erriquez:1978pg,Erriquez:1977tr,Brunner:1989kw,Fanourakis:1980si,Kuzmin:2008zz}. 
We have also compared our results with the other theoretical results in quark model obtained by 
Wu et. al.~\cite{Wu:2013kla} and Finjord et. al.~\cite{Finjord:1975zy} and in 
Cabibbo model by Kuzmin et. al.~\cite{Kuzmin:2008zz}.
We see that the most recent results in quark model by Wu et. al.~\cite{Wu:2013kla} underestimate the cross sections.
  The theoretical results presented in Fig.~\ref{fg:xsec_comp} do not include nuclear 
  medium effects while the experimental results are on nuclear targets.
Obviously experimental data with better statistics are needed to study nuclear medium effects.
 However, Cabibbo theory with $SU(3)$  symmetry seems to work well for explaining the present results on
 $ \bar \nu_\mu p \rightarrow \mu^+ \Lambda$. In Fig.~\ref{fg:xsec_p_mu_Sig}, we show the results for 
  $ \bar \nu_\mu  p \rightarrow \mu^+ \Sigma^0$ where new data are needed to draw any conclusion.
 For completeness we present the results for $ \bar \nu_\mu p \rightarrow \mu^+ \Lambda$ 
 in $ ^{40}Ar,~ ^{56}Fe$ and $^{208}Pb$ in Fig.~\ref{fg:xsec_fe_ar_pb} where we also show the effect of final state interactions.

\begin{figure}
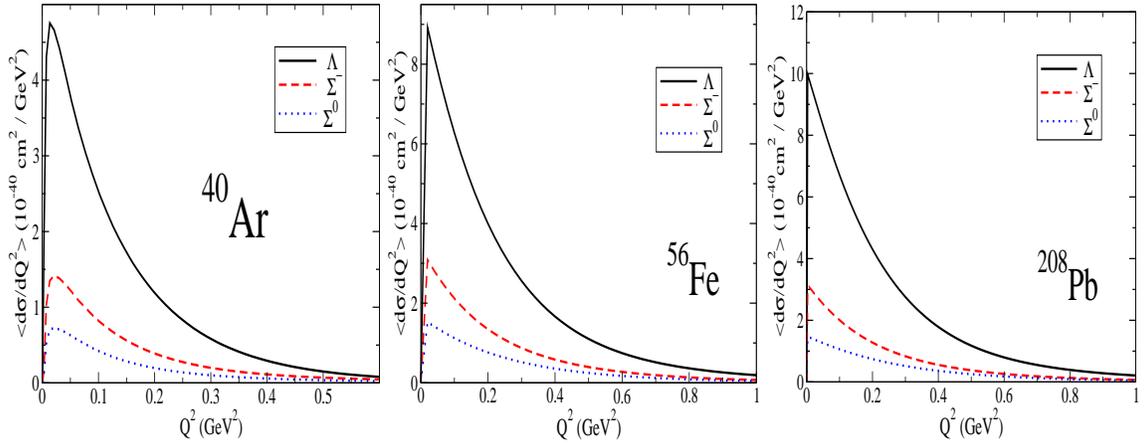
\centering
\includegraphics[height=0.26\textheight,width=0.3\textwidth]{avg_dsdq2_Ar_Microboone.eps}
\includegraphics[height=0.26\textheight,width=0.3\textwidth]{avg_dsdq2_Fe_Minerva.eps}
\includegraphics[height=0.26\textheight,width=0.3\textwidth]{avg_dsdq2_lead_minerva.eps}
\caption{$<\frac{d\sigma}{dQ^2}>$ vs $Q^2$(Left panel) in $^{12}C$, $^{56}Fe$ and $^{208}Pb$ 
nuclear targets obtained by averaging $Q^2$-distribution over the MINER$\nu$A~\cite{Fields:2013zhk} 
flux for the reaction given in Eq.~\ref{hyp-rec}. The results are presented with nuclear 
medium and final state interaction effects.}
\label{fg:q2_minerva}
\end{figure}

In Fig.~\ref{fg:q2_minerva}, we present the results for $\left \langle \frac{d \sigma}{d Q^2} \right \rangle $
in $^{40}Ar$ averaged over the MiniBooNE antineutrino spectrum~\cite{AguilarArevalo:2007it} and
for $^{56}Fe$ and $^{208}Pb$ averaged over the MINER$\nu$A  antineutrino spectrum~\cite{Fields:2013zhk}.
It should be noted that lepton energy spectrum i.e.  $\left \langle \frac{d \sigma}{d Q^2} \right \rangle $
can be easily obtained from the $Q^2$ distribution in each case. The details are given in Ref.~\cite{Alam:2014bya}.

\section{threshold production of two pions}
Two pion production is the  threshold process of inelastic reactions beyond  single pion production. 
Experimental observation of following processes on nucleon targets 
\begin{eqnarray}
 \nu_\mu p &\rightarrow& \mu^- p \pi^+ \pi^0 \\
 \nu_\mu p &\rightarrow& \mu^- n \pi^+ \pi^+ \\ 
  \nu_\mu n &\rightarrow& \mu^- p \pi^+  \pi^- 
\end{eqnarray}
have been reported at ANL~\cite{Radecky:1981fn} and BNL~\cite{Kitagaki:1990vs} experiments.
To the best of our knowledge no theoretical calculation has been reported in literature
except in the threshold region~\cite{Hernandez:2007ej,Adjei:1980nj}.
These calculations make use of an effective chiral Lagrangian to describe the interaction of 
weak currents with pions and nucleons.
It seems that the resonance contribution is as important as the 
contribution of nucleons and pions(including the contact  term) 
even in the threshold region.
It may turn out that the $2 \pi$  production is dominated by processes in which a pion  
is produced along with a $\Delta$ in the intermediate state 
  giving rise to  another pion as happens in the electroproduction~\cite{Mokeev:2008iw,Mokeev:2012vsa}.
Calculations of two pion production  in an intermediate  $\Delta$  
dominance model will be very useful in analyzing the existing data from 
ANL~\cite{Radecky:1981fn} and BNL~\cite{Kitagaki:1986ct} experiments on two pion  production and 
 data to be obtained in future from the experiments looking for neutrino oscillation.

\begin{figure}\centering
  \includegraphics[width=0.9\textwidth]{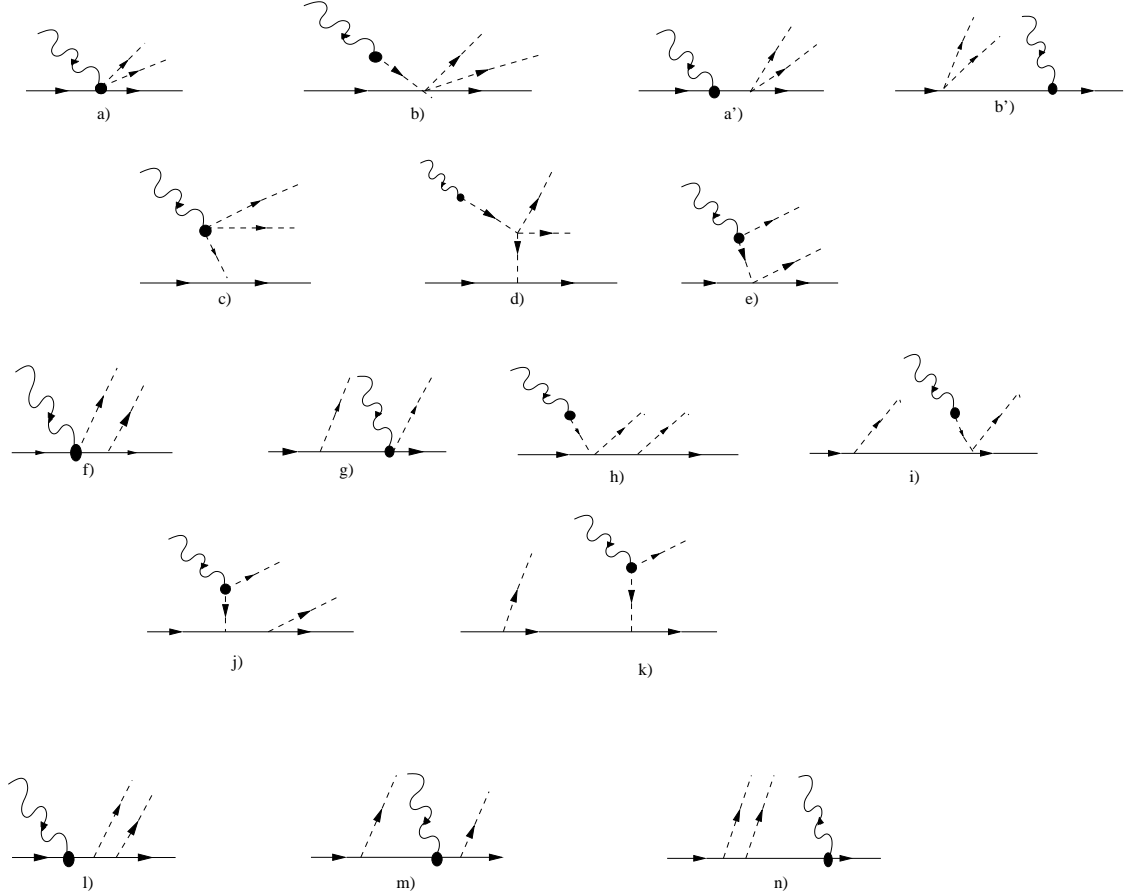}
  \caption{Nucleon pole, pion pole and contact terms contributing to
    $2\pi$ production.}
  \label{fig:fig1}
\end{figure}
In this work, we report  the calculation of  weak  charged 
current  production of two pions  in  threshold region. 
We use a nonlinear  chiral Lagrangian model which has been extensively
used to study the  single pion  production~\cite{Hernandez:2007qq}.
 The Lagrangian   describes the  interaction of pions   
 and nucleons and generates the week vector and axial vector current under a $SU(2) \times SU(2)$ 
transformation which interacts with $W^{\pm}$/$Z$ boson  as described by the Standard Model. 
The effective Lagrangian is given as 
\begin{eqnarray}
{\cal L}_ {\rm int}^\sigma&=& \frac{g_A}{f_\pi} \bar\Psi \gamma^\mu\gamma_5
\frac{\vec{\tau}}{2}(\partial_\mu \vec{\phi})\Psi
-\frac{1}{4f_\pi^2}\bar\Psi \gamma_\mu \vec{\tau}\left
(\vec{\phi}\times\partial^\mu\vec{\phi}\right)\Psi -
\frac{1}{6f_\pi^2} \left ( \vec{\phi}^{\,2}
\partial_\mu\vec{\phi}\partial^\mu \vec{\phi}-(\vec{\phi}\partial_\mu
\vec{\phi} )(\vec{\phi}\partial^\mu
\vec{\phi}) \right) + \frac{m_\pi^2}{24f_\pi^2}(\vec{\phi}^{\,2})^2
\nonumber \\
&&-\frac{g_A}{6f_\pi^3} \bar\Psi \gamma^\mu\gamma_5
\left[\vec{\phi}^{\,2} \frac{\vec{\tau}}{2}\partial_\mu \vec{\phi} -
(\vec{\phi}\partial_\mu\vec{\phi})\frac{\vec{\tau}}{2} \vec{\phi}
\right]\Psi \, ,
\label{eq:lint}
\end{eqnarray}
where $\Psi= \left (\begin{array}{c}p\cr n\end{array}\right )$ is the
nucleon field, $\vec{\phi}$ is the isovector pion field, $\vec{\tau}$ are the Pauli matrices 
and $f_\pi=93$ MeV is the pion decay constant. 
The  vector$(V_\mu)$ and axial vector$(A_\mu)$  currents 
generated by this Lagrangian under chiral transformation are given by
\begin{eqnarray}
{\vec V}^\mu &=& \underbrace{\vec{\phi} \times \partial^\mu
  \vec{\phi}}_{{\vec V}^\mu_a} + \underbrace{\bar\Psi
\gamma^\mu \frac{\vec\tau}{2} \Psi}_{{\vec V}^\mu_b} 
+ \underbrace{\frac{g_A}{2f_\pi}\bar\Psi
\gamma^\mu \gamma_5 (\vec{\phi}\times\vec{\tau})\Psi}_{{\vec V}^\mu_c}
  \overbrace{-
\frac{1}{4f_\pi^2} \bar\Psi \gamma^\mu\left [
  \vec{\tau}\vec{\phi}^{\,2}-
  \vec{\phi}(\vec{\tau}\cdot\vec{\phi})\right]\Psi -
\frac{\vec{\phi}^{\,2}}{3f_\pi^2}(\vec{\phi}\times \partial^\mu
\vec{\phi})}^{{\vec V}^\mu_d} + {\cal O}(\frac{1}{f_\pi^3})
  \label{eq:vcurrent} \\ 
{\vec A}^\mu &=& \underbrace{f_\pi \partial^\mu
  \vec{\phi}}_{{\vec A}^\mu_a} + \underbrace{g_A\bar\Psi
\gamma^\mu \gamma_5\frac{\vec\tau}{2} \Psi}_{{\vec A}^\mu_b} 
+ \underbrace{\frac{1}{2f_\pi}\bar\Psi
\gamma^\mu (\vec{\phi}\times\vec{\tau})\Psi}_{{\vec A}^\mu_c}\nonumber \\ 
 &+& 
 \overbrace{
\frac{2}{3f_\pi}\left[\vec{\phi} (\vec{\phi}\cdot\partial^\mu\vec{\phi})-
\vec{\phi}^{\,2}\partial^\mu\vec{\phi}\,\right]
-\frac{g_A}{4f_\pi^2} \bar\Psi \gamma^\mu\gamma_5\left [
  \vec{\tau}\vec{\phi}^{\,2}-
  \vec{\phi}(\vec{\tau}\cdot\vec{\phi})\right]\Psi }^{{\vec
  A}^\mu_d} + {\cal O}(\frac{1}{f_\pi^3}) \label{eq:acurrent}
\end{eqnarray}
 These currents couple to $W^\pm (Z)$ boson
for charged current( neutral current)  interaction with nucleon and pion  as described in the Standard Model.
 Various terms in $V_\mu$ and $A_\mu$  describe the vector and 
 axial vector couplings for $WNN$, $WN\pi$ $WNN\pi$ etc. vertices.
 And the matrix elements for various Feynman diagrams can be 
 calculated for $\bar \nu N \rightarrow \mu^- \pi \pi N$ process using the rules of covariant perturbation theory.
 Using these  currents given in Eqs.~\ref{eq:vcurrent} and~\ref{eq:acurrent} 
 we obtain following 16 diagrams for $2\pi$ productions shown in Fig.~\ref{fig:fig1},
 for which the matrix element can be explicitly written using 
 $W^+ + N \rightarrow N$, $W^+ + N \rightarrow N \pi$, 
$W^+ + N \rightarrow N \pi \pi $ and $ W^+ \pi \rightarrow \pi $ 
vertices from Eqs.~\ref{eq:vcurrent} and~\ref{eq:acurrent}.
The $ \pi NN$  and $\pi \pi NN$ etc. couplings are derived from Eq.~\ref{eq:lint}. 
With these matrix elements for the hadronic transition current $J_\mu$, the matrix element for the process 
$\nu_\mu (k) + N(p) \rightarrow \mu^-( k^\prime) + \pi( k_{\pi_1} ) + \pi( k_{\pi_2} ) + N(p^\prime) $ is written as 
\begin{equation}
 {\cal M} = \frac{G_F}{\sqrt 2} \cos \theta_c l^\mu J_\mu ,
\end{equation}
where 
\begin{equation}
 l^\mu = \bar u (k^\prime) \gamma^\mu (1-\gamma^5) u(k)
\end{equation}
and 
\begin{equation}
 J_\mu = \left \langle N^*(p^\prime) \pi( k_{\pi_1} ) \pi( k_{\pi_2} ) | j_\mu | N(p) \right \rangle 
\end{equation}

The present formalism gives the correct form of the nucleon vector ($V^\mu_b$) and axial vector ($A^\mu_b$) 
currents which couple to the $W^\pm$ boson but go with a point coupling for the $WNN$ vertex. 
The form factors are phenomenologically introduced at this vertex which are
consistent with electron proton and neutrino nucleon scattering. 
The following forms are used for the matrix element for $V^\mu_b$ and  $A^\mu_b$ currents
\begin{equation}
V_b^\mu(q) = 2\times\left(f_1^V(q^2)\gamma^\mu + {\rm
i}\mu_V \frac{f_2^V(q^2)}{2M}\sigma^{\mu\nu}q_\nu\right), \qquad
A_b^\mu (q)=  g_1(q^2) \times \left(\gamma^\mu\gamma_5 + 
\frac{\slashchar{q}}{m_\pi^2-q^2}q^\mu\gamma_5 \right) \, .
\label{eq:axial1} 
\end{equation}
with $f_{1,2}^V (q^2) = f_{1,2}^p (q^2) - f_{1,2}^n (q^2)$ where the vector form factors 
$f_{1,2}^{p, n}$ and the axial vector form factor $g_1 (q^2)$ are given by Eq.~\ref{Vec_FF}. 

The explicit expressions for the matrix elements of 16 Feynman diagrams shown in Fig.~\ref{fig:fig1} 
contributing to $J_\mu$ are given in Ref.~\cite{Hernandez:2007ej}. 
The cross section is then expressed as 
\begin{equation}
\frac{d\sigma_{\nu_ll}}{d\Omega(\hat{k^\prime})dE^\prime} = 
 \frac{G^2}{4\pi^2}\frac{|\vec{k}^\prime|}{|\vec{k}|}
  L_{\mu\sigma}\left(W^{\mu\sigma}_{{\rm CC}2\pi}\right)
\label{eq:sec}
\end{equation}
with 
\begin{eqnarray}\label{eq:lep} 
L_{\mu\sigma} &=& (L_s)_{\mu\sigma} \pm i
 (L_a)_{\mu\sigma} =
 k^\prime_\mu k_\sigma +k^\prime_\sigma k_\mu
- g_{\mu\sigma} k\cdot k^\prime \pm 
i\epsilon_{\mu\sigma\alpha\beta}k^{\prime\alpha}k^\beta \\ 
W^{\mu\sigma}_{{\rm CC} 2\pi} &=& 
 \overline{\sum_{\rm spins}} \int
\frac{d^3p^\prime}{(2\pi)^3} \frac{M}{E^\prime_N}
\frac{d^3k_{\pi_1}}{(2\pi)^3} \frac{1}{2E_{\pi_1}}
\frac{d^3k_{\pi_2}}{(2\pi)^3} \frac{1}{2E_{\pi_2}}
(2\pi)^3 \delta^4(p^\prime + k_{\pi_1} + k_{\pi_2}  - q - p)  \nonumber \\ &&
\langle N^\prime\pi_1\pi_2|j^\mu_{\rm cc+}(0)|N\rangle
\langle N^\prime\pi_1\pi_2|j^\sigma_{\rm cc+}(0)|N\rangle^*
\end{eqnarray}

\section{Contribution of $N^*(1440)$ Resonance}

The Roper $N^*(1440)$ is the lowest lying resonance with significant coupling to two pion
decay mode in $S$ state and is expected to contribute in the threshold region. 
The $N^* \rightarrow N \pi \pi $ coupling has been studied in $\pi N \rightarrow \pi \pi N$ 
and $ NN \rightarrow NN \pi \pi $ reactions~\cite{AlvarezRuso:1997mx} where $N^*$ contribution found to be important. 
The $N^* \rightarrow N \pi \pi $ is described by the Lagrangian
\begin{equation}
  {\cal L}_{N^*N\pi\pi} = 
  -c_1^*\frac{m_\pi^2}{f_\pi^2}\bar{\psi}_{N^*}\vec{\phi}^2\Psi +
   c_2^*\frac{1}{f_\pi^2}\bar{\psi}_{N^*}(\vec{\tau}\partial_0\vec{\phi})
   (\vec{\tau}\partial_0\vec{\phi})\Psi + h.c.\, ,
\label{eq:pipilag}
\end{equation}
with $c_1^*=-7.27$ GeV$^{-1}$, $c_2^*=0$ GeV$^{-1}$~\cite{AlvarezRuso:1997mx}.

\begin{figure}\centering
  \includegraphics[width=0.9\textwidth]{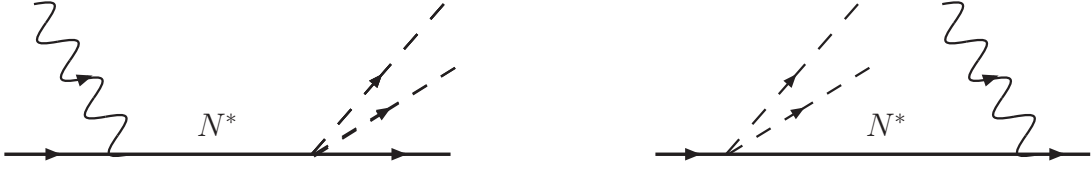}
  \caption{Direct  (left) and crossed  (right) Roper excitation  contributions to  $2\pi$ production.}
  \label{fig:rop}
\end{figure}
The Feynman diagram for $\nu_\mu N \rightarrow \mu^- N \pi \pi $ 
through $N^*$ excitation is shown in Fig.~\ref{fig:rop}, for which 
the matrix element is written as
\begin{equation}
 {\cal M} = 2 g^* \bar u (p^\prime) \left[ S_R (p+q) J^\mu_{R}(q) + \tilde J^\mu_{R} (q) S_R(p^\prime - q) \right] u(p) 
 \;\; ; \qquad \tilde J^\mu_{R} = \gamma^0 {J^\mu_{R}}^\dagger \gamma^0 
\end{equation}
where $S_R$ is the propagator for Roper resonance  
\begin{equation}
S_R(p_*) = \frac{\slashchar{p}_*+M_R}{p_*^2-M_R^2 + 
i(M_R+W)\Gamma_{\rm tot}(W)/2} \, .
\label{eq:roprop}
\end{equation}
and
\begin{equation}\label{eq:mat_roper}
J^\mu_{R} = 
\frac{F_1^{V*}(q^2)}{\mu^2}(q^\mu\slashchar{q}-q^2\gamma^\mu)
+ i\frac{F_2^{V*}(q^2)}{\mu}\sigma^{\mu\nu}q_\nu 
- G_A\gamma^\mu\gamma_5 - \frac{G_P}{\mu}q^\mu\slashchar q\gamma_5 
- \frac{G_T}{\mu} \sigma^{\mu\nu}q_\nu\gamma_5
\end{equation}
with $\mu = M_N + M_R$; $M_N$ and $M_R$ are respectively the mass of nucleon and Roper resonance 
and $g^* = - c_1^* \frac{m_\pi^2}{f_\pi^2}$. 

The axial vector form factor $G_P$ is obtained using PCAC hypothesis as 

\begin{equation}
 G_P(q^2)  =\frac{M_N + M_R}{m_\pi^2-q^2}G_A(q^2) \, .
\label{eq:pcac}
\end{equation}
and $G_{A}(0)$ is given by Goldberger-Treiman relation 
\begin{equation}
G_A(0)= 2f_\pi \frac{\tilde{f}}{m_\pi}=0.63
\label{eq:gtnn*}
\end{equation}
with $\tilde{f}$ is strength of $N^{\star}N\pi$ coupling determined by 
the $N^{\star} \rightarrow N \pi$ decay width $\Gamma$
\begin{equation}
\Gamma=\frac{3}{2\pi}\left(\frac{\tilde{f}}{m_\pi}\right)^2 
\frac{M}{W}|q_{\rm cm}|^3 \, ,
\label{eq:ropwidth}
\end{equation}
with $W$ as center of mass energy of $\pi N$ state and $|{\bf q}_{\rm cm}|$ is the
center of mass momentum. With $\Gamma$ = 350 MeV, we obtain $\tilde{f}$ = 0.48.
For $G_A(q^2)$, $q^2$ dependence is assumed to be of dipole form i.e. 
\begin{equation}
  G_A(q^2) = \frac{G_A(0)}{(1-q^2/M_{A}^2)^2} \, ,
\label{eq:axcoup}
\end{equation}
with $M_{A} = 1$ GeV.
 The isovector vector form factors $F_{1}^{V*}=F_{1p}^{*} - F_{1n}^{*}$ and
 $F_{2}^{V*}=F_{2p}^{*} - F_{2n}^{*}$ are determined from the helicity amplitudes defined as

\begin{equation}
A_{1/2}^N = \sqrt{\frac{2\pi\alpha}{k_R}}\langle N^*\uparrow|
\sum_{{\rm pol}}
\epsilon\cdot 
j_{{\rm e.m.}}(0) |N\downarrow \rangle\,\xi \label{eq:heli0}\\
S_{1/2}^N = \sqrt{\frac{2\pi\alpha}{k_R}}\frac{|\vec{q}|}{\sqrt{-q^2}}
\langle N^*\uparrow|
\sum_{{\rm pol}}\epsilon\cdot j_{{\rm e.m.}}(0) |N\uparrow\rangle\,\xi \, ,
\label{eq:heli}
\end{equation}
where the polarization vectors are given by
\begin{equation}
\epsilon^{\pm}=\frac{1}{\sqrt{2}}(0,\mp 1,-i,0)\, ,
\label{eq:heli2}\end{equation}
and for a photon of momentum $q$ moving along the positive z-axis
\begin{equation}
\epsilon^{0}=\frac{1}{\sqrt{-q^2}}(|\vec{q}|,0,0,q^0)\, .
\label{eq:heli3}\end{equation}
with $N$ for proton or neutron, $\alpha=1/137$, $q$ is the
 momentum of the virtual photon, $k_R=(W^2-M^2)/2W$ and $W$ is the center of mass 
 energy of the Roper.

The EM  $\gamma N\to N^*$ current is written as
\begin{equation}
\langle N^*;\vec p_{*}=\vec p+\vec q\,|j_{{\rm e.m.}}^\alpha(0)|N;\vec p\,\rangle = \bar{u}_*(\vec p_{*})
\left[\frac{F_1^{N*}(q^2)}{\mu^2}(q^\alpha\slashchar{q}-q^2\gamma^\alpha)
+ i\frac{F_2^{N*}(q^2)}{\mu}\sigma^{\alpha\nu}q_\nu\right] u(\vec p\,)\, .
\label{eq:heli4}
\end{equation}
Using Eqs.~\ref{eq:heli}-\ref{eq:heli4}, we obtain the following relations:
\begin{eqnarray}
A_{1/2}^N &=& |\vec{q}| g(q^2)\left[\frac{F_2^{N*}}{\mu}-\frac{q^2}{W+M} 
\frac{F_1^{N*}} {\mu^2}\right] \label{eq:heli5a}  \\
S_{1/2}^N &=& \frac{1}{\sqrt{2}}|\vec{q}|^2 g(q^2)
\left[\frac{F_1^{N*}}{\mu^2}-\frac{F_2^{N*}}{\mu} 
\frac{1}{W+M}  \right] \, ,
\label{eq:heli5b}
\end{eqnarray}
with
\begin{equation}
  g(q^2)=\sqrt{\frac{8\pi\alpha(W+M)W^2}{M(W-M)((W+M)^2-q^2)}} \, .
\end{equation}
Inverting Eqs.~\ref{eq:heli5a} and \ref{eq:heli5b} to obtain $F_1^{N*}$ and $F_2^{N*}$
in terms of $A^{N}_{1/2}$ and $S^{N}_{1/2}$ and using quark model predictions of 
$A^{n}_{1/2}=-2/3 A^p_{1/2}$ and
$S^{n}_{1/2}=0$ we obtain
\begin{eqnarray}
F_1^{V*}&=&\frac{F_1^{p*}((M+W)^2-5 q^2/3)+2/3 F_2^{p*}(M+W) \mu}{(M+W)^2-q^2}\\
F_2^{V*}&=&\frac{F_2^{p*}(5(M+W)^2-3 q^2)\mu-2 F_1^{p*}q^2(M+W)}{3((M+W)^2-q^2)\mu}\, .
\end{eqnarray}
The $q^2$ dependence of $F_1^{p*}$ and $F_2^{p*}$ have been obtained by
fitting the data on proton--Roper  electromagnetic 
transition form factors to the experimental results in helicity  amplitudes
and their forms are obtained as
\begin{equation}
F_1^{p*}(q^2) = \frac{g_1^p/D_V}{1-q^2/X_1M_V^2} \, \qquad 
F_2^{p*}(q^2) = \frac{g_2^p}{D_V}\left(1 - X_2\ln\left(1-\frac{q^2}{1\,{\rm GeV}^2}\right) \right)
\end{equation}

\section{Results}
We present numerical results for the different channels using the matrix elements corresponding to various Feynman diagrams
shown in Fig.~\ref{fig:fig1} and Fig.~\ref{fig:rop}.
 All the coupling constants and form factors are fixed as described in the text
 through Eqs~\ref{eq:lint}-\ref{eq:acurrent}.
  The relative phases  between different diagrams in Fig.~\ref{fig:fig1} are fixed by the Lagrangian itself.
  
In the case of resonance contribution the sign of $N^* N \pi$  coupling is taken to be same as $N N \pi$  coupling.
The axial  form factors are used as given in Eqs~\ref{eq:mat_roper}-\ref{eq:gtnn*}.  For vector  form factors 
$F_1^{V^*}$ and $F_2^{V^*}$ we have considered four cases  i.e. FF1, FF2, FF3, FF4 as follows.
\begin{enumerate}
 \item[FF1:] Our fit of helicity amplitude given by MAID~\cite{Hernandez:2007ej}
 \item[FF2:] Form Factors as determined in the quark model of Meyer et al.~\cite{Meyer:2001js}
 \item[FF3:] Parameters given by Lalakulich et al. \cite{Lalakulich:2006sw}
 \item[FF4:] MAID analysis of helicity amplitudes~\cite{Drechsel:2007if}
\end{enumerate}

\begin{figure}
 \includegraphics[width=0.6\textwidth]{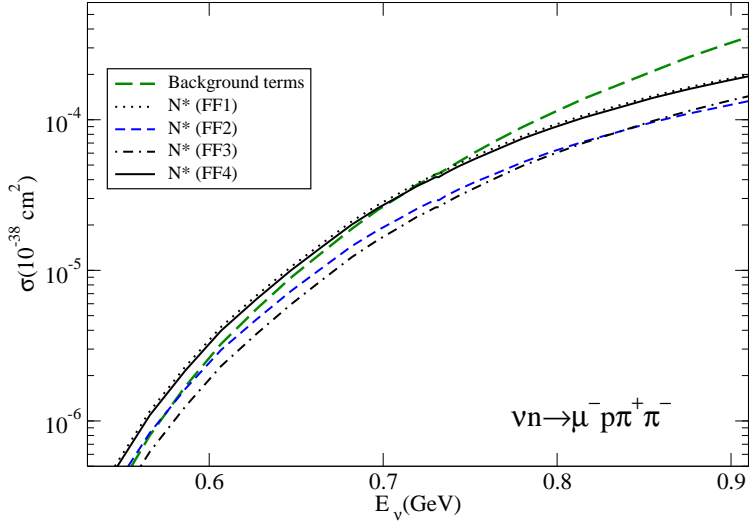}
\caption{Cross section for the $\nu_\mu n\to \mu^- p \pi^+\pi^-$ reaction as a function of the neutrino energy. The 
interference between background and the $N^*$ contribution is not shown. See text for details.}
\label{fig:3}     
\end{figure}

\begin{figure}
 \includegraphics[width=0.6\textwidth]{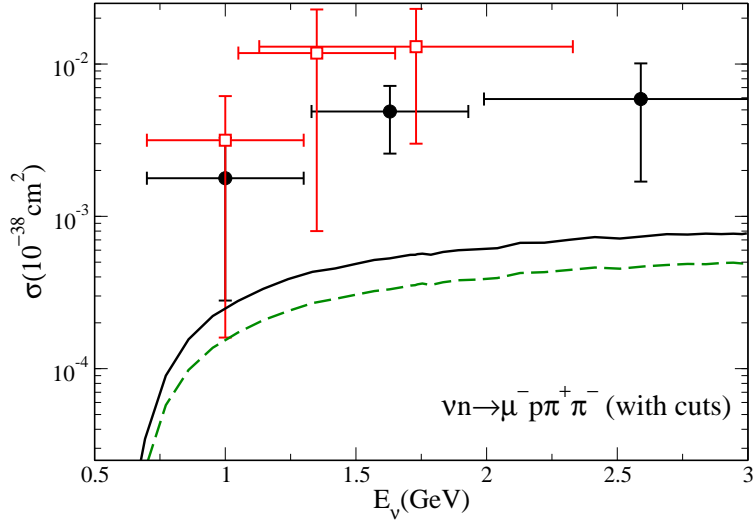}
\caption{Cross section for the $\nu_\mu n\to \mu^- p \pi^+\pi^-$ with cuts
as explained in the text.  Dashed line: Background terms. Solid line:
Full model with set 1 of nucleon-Roper transition FF.  Data are from
Ref.~\cite{Kitagaki:1986ct} (solid circles) and Ref.~\cite{Day:1984nf}
(open squares).}
\label{fig:5}     
\end{figure}

In Fig.~\ref{fig:3}, we present the results for the cross section for  
the process $\nu_\mu n \rightarrow \mu^- p \pi^+ \pi^-$.
We show separately the contribution of background terms(pion and nucleon, 
including the contact terms) and Roper resonance.
We see that at lower energies the contribution of Roper is comparable to the
background  terms and can be more or less than background contribution depending upon the form factors used.
 The resonance contribution is sensitive to $F_2^{V^*}(q^2)$. 
 At higher energies $E_\nu > 0.7 GeV$, background terms dominate.
 In Fig.~\ref{fig:5}, we present the results for this process with experimental
 results~\cite{Kitagaki:1986ct, Day:1984nf} in kinematical region  restricted by
\begin{eqnarray}
q_\pi^2 &\leq& \left((1 + \eta/2)m_\pi\right)^2 \, ,\\
p\cdot q_\pi &\leq& (M + (1+\eta)m_\pi)^2 - M^2 -m_\pi^2 \,  \\
p^\prime\cdot q_\pi &\leq& (M + (1+\eta)m_\pi)^2 - M^2 -m_\pi^2 \, ,
\end{eqnarray}
Obviously our model underestimates the experimental results indicating the contribution of other  resonances
or even the presence of 
 intermediate $\Delta \pi$  state  in this  kinematics region.

\begin{figure}
  \includegraphics[width=0.6\textwidth]{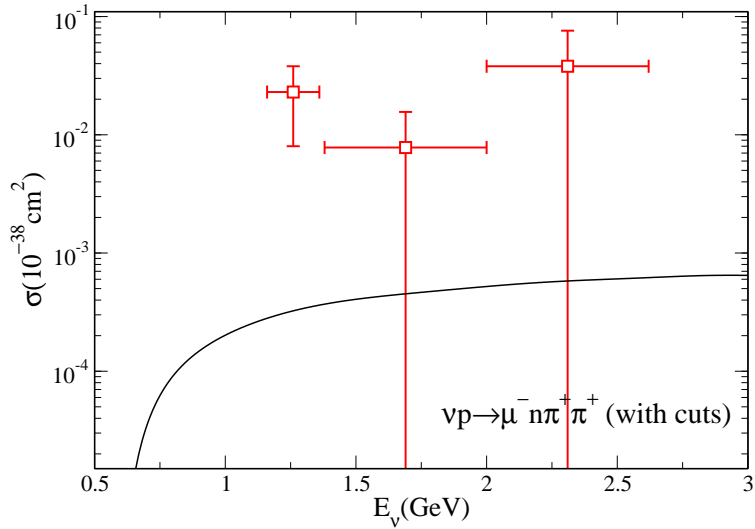}
  \caption{Cross section for the $\nu_\mu p\to \mu^- n \pi^+\pi^+$ with cuts as 
explained in the text. Note that there are no contributions from 
the $N^*(1440)$ resonance to this channel. Data are from Ref.~\cite{Day:1984nf}.}
\label{fig:6}     
\end{figure}

\begin{figure}
\includegraphics[width=0.6\textwidth]{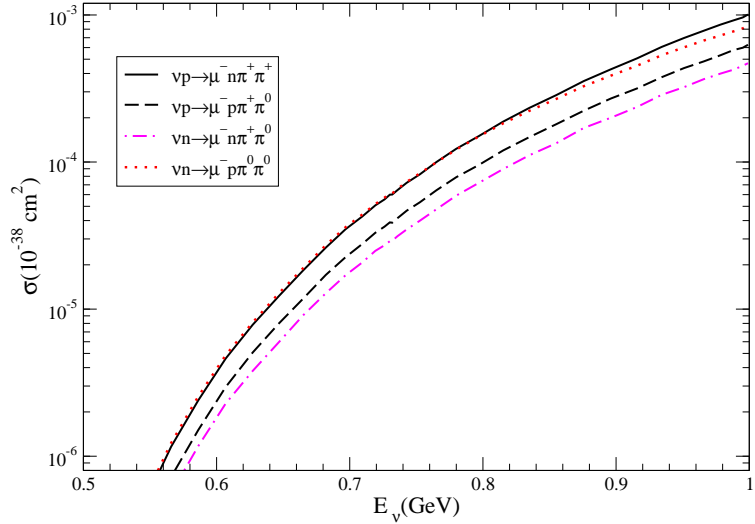}
\caption{Cross sections as a function of the neutrino energy.  All
calculations correspond to the full model with the FF1 set of nucleon-Roper
transition form factors.}
  \label{fig:7}     
\end{figure}

In Fig.~\ref{fig:6}, we  present the results for $\nu_\mu p \rightarrow \mu^- n \pi^+ \pi^+ $  
and compared them with the  experimental results~\cite{Day:1984nf}.
Here again, our results are lower than the experimental results in this channel. 
It should be noted that while our results underestimate the experimental results 
in all these channels, they are larger than the results obtained by Adjei et. al.~\cite{Adjei:1980nj}.
In view of this an improved calculation is highly desired for weak pion production 
in threshold region as well as at higher neutrino energies. 
 In Fig.~\ref{fig:7} we present our  prediction for the cross section for other channels like:
 \begin{eqnarray}
  \nu_\mu p &\to& \mu^- p \pi^+\pi^0 \\ 
  \nu_\mu p  &\to& \mu^- n \pi^+\pi^+ \\ 
  \nu_\mu n &\to& \mu^- n \pi^+\pi^0 \\
  \nu_\mu n &\to& \mu^- p \pi^0\pi^0
 \end{eqnarray} 
 
 \section{Conclusion}
 In this work we have presented a review of the   theoretical calculation of 
 reactions for the production of leptons due 
 to quasielastic production of hyperons $\bar \nu_\mu   N \rightarrow \mu^+ Y$ and two pion production 
$ \nu_\mu    N \rightarrow \mu^- \pi \pi^\prime N$.
These are the lowest  threshold processes beyond the single pion  production and may be seen in the present
generation of neutrino oscillation experiments some of them which are producing results on cross section  measurements.
 Phenomenological  cross section have been used for these processes in modeling the neutrino cross section for 
 validating the Monte Carlo generators for neutrino events.
 It is desirable that a theoretical calculation of the cross section for these inelastic 
 processes is made available for modelings the neutrino-nucleus cross section in this energy region.

 A theoretical understanding of quasielastic hyperon production processes will help us to understand the role of 
 symmetries of weak interaction currents at higher momentum transfer in the quasielastic production of hyperons.
 In the case of two pion  production,  there is an urgent need of going beyond the  region of threshold
 production  to understand the reaction dynamics of two pion  production in weak  interaction.
 Even in the threshold region,  there is a need to include other contributions  not considered in this work.
 Experimentally,  efforts should be made to observe these reactions in the ongoing experiments
 looking for neutrino oscillation in the few GeV energy region.

\begin{theacknowledgments}
One of the authors(S. K. Singh) is thankful to the organizers of the CETUP workshop and specially to Profs. 
Barbara Szczerbinska, Jan Sobczyk, L. Alvarez-Ruso for the warm hospitality and financial support.
\end{theacknowledgments}

 \bibliographystyle{aipproc}   

\end{document}